\newcommand{\be}{\begin{equation}}\newcommand{\ee}{\end{equation}}
\newcommand{\bea}{\begin{eqnarray}}\newcommand{\eea}{\end{eqnarray}}
\newcommand{\nn}{\nonumber\\}\newcommand{\p}[1]{(\ref{#1})}

\newcommand{\cD}{{\cal D}}

\newcommand{\hK}{{\hat K}}
\newcommand{\hD}{{\hat D}}
\newcommand{\bQ}{{\overline Q}}
\newcommand{\bS}{{\overline S}}
\newcommand{\bD}{{\overline D}}
\newcommand{\hS}{{\hat S}}
\newcommand{\hbS}{{\hat{\overline S}}}
\newcommand{\bt}{{\bar\theta}}
\newcommand{\bpsi}{{\bar\psi}}
\newcommand{\bxi}{{\bar\xi}}
\documentstyle[12pt]{article}

\begin{document}

\begin{titlepage}
\begin{flushright}
hep-th/0305255 \\
May, 2003
\end{flushright}

\begin{center}
{\Large \bf{`Conformal Theories - AdS Branes' Transform, \vspace{0.2cm} \\
or One More Face of AdS/CFT
}}
\vspace{1.5cm}\\
E. Ivanov \vspace{1cm} \\
\centerline{\it Bogoliubov  Laboratory of Theoretical Physics, JINR,}
\centerline{\it 141 980 Dubna, Moscow Region, Russia} \vspace{0.3cm}
{\bf Abstract}
\end{center}
\noindent The AdS/CFT transformation relates two nonlinear realizations
of (super)conformal groups: their realization
in the appropriate field theories in Minkowski space with a Goldstone dilaton field and
their realization as (super)isometry groups of AdS (super)spaces. It exists already at the
classical level and maps the field variables and space-time coordinates of
the given (super)conformal field theory in $d$-dimensional Minkowski space ${\cal M}_d$
on the variables of a scalar codimension one (super)brane in AdS$_{d+1}$
in a static gauge, the dilaton being mapped on the transverse AdS brane coordinate.
We explain the origin of this coordinate map and describe some its implications, in particular,
in $d=1$ models of conformal and superconformal mechanics. We also give a suggestive
geometric interpretation of this AdS/CFT transform in the pure bosonic case
in the framework of an extended $2d+1$-dimensional conformal space involving
extra coordinates associated with the generators of dilatations and conformal boosts.

\vskip3cm
\begin{center}
{\it Submitted to Proceedings of the Seminar ``Classical and Quantum
Integrable Systems'' dedicated to the memory of M.V. Saveliev (Protvino,
Russia, 8--11 January 2003)}
\end{center}

\end{titlepage}

\section{Introduction}
The cornerstone of AdS/CFT correspondence \cite{mald}-\cite{wit}
is the assertion that the isometry group of an AdS$_n\times S^m$ background
in which some type IIB string theory lives is identical to the
standard conformal group (times the group of internal $R$ symmetry) of the
appropriate conformal field theory living on the $(n-1)$-dimensional
Minkowski space interpreted as a boundary of AdS$_n\,$. The supersymmetric
version of this correspondence deals with the appropriate realizations of
superconformal groups. Originally, the AdS/CFT correspondence was formulated as
a duality between IIB string theory compactified on AdS$_5\times S^5$ and
$N=4$ super Yang-Mills theory in $d=4$ Minkowski space.

It was shown in \cite{5}-\cite{solv}, \cite{mald} that the invariance group of the
worldvolume action of some probe brane in an AdS$_n\times S^m$ background
(e.g. D3-brane in AdS$_5\times S^5$)
can be realized as a field-dependent modification
of the standard (super)conformal transformations of the worldvolume.
On the other hand, in the spirit of
the AdS/CFT correspondence, the AdS
superbrane worldvolume actions are expected to appear as the result of summing up
some leading (in external momenta) terms in the low-energy
quantum effective actions of the corresponding Minkowski space (super)conformal field
theories in a phase with spontaneously broken (super)conformal symmetry
(e.g. the effective action of
$N=4$ SYM theory in Coulomb branch) \cite{mald}, \cite{chep}-\cite{ats}.
It was argued in \cite{14,13} that the modified (super)conformal
transformations can be understood as a quantum deformation of the standard
(super)conformal transformations.

In this contribution, basically following the line of refs. \cite{0,00},
we expound a different viewpoint on the interplay between
the standard and modified (super)conformal transformations. The basic statement is that
any conformal field theory in $d=p+1$-dimensional Minkowski space
in a phase with spontaneously
broken conformal symmetry and so with the dilaton Goldstone field can be brought,
by an invertible change of variables, into
the form in which it respects invariance just under the above mentioned field-dependent
conformal transformations. Using this relation between the conformal
and AdS bases (AdS/CFT map or transform), one can rewrite any conformal field theory with
a dilaton among the involved fields in terms of the variables of
the corresponding scalar AdS brane in a static gauge, and vice versa. The AdS images of the
minimal conformally-invariant Lagrangians (i.e. those containing terms with no more
than two derivatives) prove to necessarily include non-minimal terms composed of the
first extrinsic curvature of the brane. On the other hand,
the conformal field theory image
of the minimal brane Nambu-Goto action is a non-polynomial and higher-derivative
extension of the minimal Minkowski space conformal actions. We also discuss
some further implications of the AdS/CFT transformation and its supersymmetric
extension in the quantum-mechanical $d=1$ systems \cite{00,000}.
Also, a novel geometric interpretation of this transformation
will be given. It highlights the relevance of some extended coset manifolds of
(super)conformal groups as ambient manifolds for the above sort of the
`conformal theories - AdS branes' correspondence. In the bosonic case
it is the $(2d+1)$-dimensional coset manifold $CM^{2d+1} \sim SO(2,d)/SO(1,d-1)$.
It includes, besides the standard Minkowski coordinate, also independent coordinates
associated with the generators of dilatations and conformal boosts.

\section{Two nonlinear realizations of conformal group in ${\cal M}_d$}
The group-theoretical origin of the AdS/CFT map to be discussed takes root
in the existence of two different nonlinear realizations of the conformal
group in $d$ dimensions.

The algebra of conformal group $SO(2,d)$ of $d = p+1$-dimensional Minkowski space
reads
\bea\label{confbasis}
&& \left[ M_{\mu\nu},M^{\rho\sigma}\right] = 2 \delta^{[\rho}_{[\mu}M_{\nu]}{}^{\sigma]}\; ,\;
\left[ P_{\mu},M_{\nu\rho}\right] =-\eta_{\mu [ \nu}P_{\rho]}\;,\;
\left[ K_{\mu},M_{\nu\rho}\right] =-\eta_{\mu [ \nu}K_{\rho]}\;, \nn
&& \left[ P_{\mu},K_{\nu}\right] =2\left( -\eta_{\mu\nu} D +2 M_{\mu\nu} \right)\;,\;
\left[ D, P_{\mu}\right] = P_{\mu}\;, \; \left[ D, K_{\mu}\right] =
-K_{\mu}\;,
\eea
where we antisymmetrize with the factor $1/2$. The standard nonlinear realization of
this group (see e.g. \cite{cfstand}) is defined as left shifts of an element of the coset
$SO(2,d)/SO(1,d-1)$:
\be\label{confcoset}
g=e^{y^{\mu}P_{\mu}}e^{\varphi D} e^{\Omega^\mu K_\mu} \; .
\ee
The shifts with the parameters $a^\mu, b^\mu$ and $c$ related to the generators
$P_\mu, K_\mu$ and $D$ induce the familiar conformal transformations of the coset
coordinates
\be
\delta y^\mu = a^\mu + c\,y^\mu + 2\,(yb)y^\mu - y^2\,b^\mu~, \; \delta \varphi = c
+2\,yb~.
\label{trstand}
\ee
The left-covariant Cartan 1-forms are defined as follows
\bea\label{confcartan}
g^{-1}dg &=& \omega_P^\mu\,P_\mu + \omega_D\, D +
\omega_M^{\mu\nu}\,M_{\mu\nu} + \omega_K^\mu\,K_\mu \nn
&=&e^{-\varphi}dy^{\mu}P_{\mu} +
\left( d\varphi -2e^{-\varphi}\Omega_{\mu}dy^{\mu}\right)D-
   4e^{-\varphi}\Omega^{\mu}dy^{\nu} M_{\mu\nu}  \nn
&& + \left[ d\Omega^{\mu}-\Omega^{\mu}d\varphi
+e^{-\varphi}\left(2\Omega_\nu dy^\nu \Omega^\mu -
   \Omega^2 dy^\mu\right)\right] K_\mu \; .
\eea
The vector Goldstone field $\Omega^\mu(x)$ can be covariantly expressed
through the dilaton $\varphi(x)$ \cite{IH}
\bea
\omega_D=0 \;\Rightarrow\; \Omega_\mu = {1\over 2}\,e^{\varphi} \partial_{\mu}
\varphi~, \; \omega_P^{\mu} = e^{-\varphi}dy^{\mu}~, \; \omega_K^{\mu}=
 d\Omega^{\mu}- e^{-\varphi}\Omega^2 dy^\mu \;. \label{confih}
\eea
The covariant derivative of $\Omega^\mu$ is defined by the relation
\bea
\omega^\mu_K = \omega^\nu_P{\cal D}_\nu\Omega^\mu \; \Rightarrow \;
{\cal D}_\nu\Omega^\mu
= {1\over 2}
\,e^{2\varphi}\left[\partial_\nu\partial^\mu\varphi +
\partial_\nu\varphi \partial^\mu \varphi
-{1\over 2}\left(\partial \varphi\partial\varphi\right)\delta^\mu_\nu \right].\label{covOm}
\eea
The conformally invariant measure of integration over $\{y^\mu \}$ is defined as
the exterior product of $d$ 1-forms $\omega^\mu_P$
\be
S_1 = \int \mu(y) = \int d^{(p+1)} y\, e^{-(p+1) \varphi}~.\label{S1}
\ee
The covariant kinetic term of $\varphi$ can be constructed as
\be
S_\varphi^{kin} = \int d^{(p+1)} y\, e^{-(p+1) \varphi}\,{\cal D}_\mu\Omega^\mu =
{1\over 4}(p-1)\int d^{(p+1)}y
\, e^{(1-p)\varphi}\,\partial \varphi\partial\varphi\;. \label{Skin1}
\ee

In any field theory with spontaneously broken conformal symmetry, it is
always possible to make a field redefinition which splits the full set of
scalar fields of the theory into the dilaton $\varphi$ with the transformation
law \p{trstand} and the subset of fields which are scalars of weight zero
under conformal transformations. In this sense, the above nonlinear
realization is universal.

While the standard nonlinear realization of $SO(2,d)$ describes a spontaneously
broken phase of conformally invariant field theories, there is another sort of
nonlinear realizations of the same group \cite{dik} which proves to be relevant
to the description of
codimension one branes on AdS$_{d+1}$. In this realization, $SO(2,d)$ acts as the
group of motion of AdS$_{d+1}$. It is related to the existence of the so-called
AdS basis in the algebra \p{confbasis}.

In the AdS basis, we introduce the following generators
\be\label{adsgenerators}
\hK_{\mu} =mK_{\mu}-\frac{1}{2m}P_{\mu}\;,\; \hD=mD \;,
\ee
where $m$ is the inverse AdS radius. The basic relations of the $SO(2,d)$ algebra
become
\bea\label{adsbasis}
\left[ \hK_{\mu},\hK_{\nu}\right] = 4 M_{\nu\mu}\,,
\left[ P_{\mu},\hK_{\nu}\right] = 4m M_{\mu\nu} - 2\eta_{\mu\nu} \hD\,,
\left[ \hD, P_{\mu}\right] = mP_{\mu}\,, \left[\hK_{\mu}, \hD\right] =
P_{\mu} +m\hK_{\mu}.
\eea
The main difference between \p{adsbasis} and \p{confbasis} is that the generators
$(\hat{K}^\mu, M_{\rho\nu})$ generate the semi-simple subgroup $SO(1,d)$ of $SO(2,d)$,
while the subgroup with $(K^\mu, M_{\rho\nu})$ has the structure of a semi-direct product.
As a result, in the coset element \p{confcoset} rewritten in the new basis
\be\label{adscoset}
g=e^{x^{\mu}P_{\mu}}e^{q \hD} e^{\Lambda^\mu \hK_\mu}\,,
\ee
$x^\mu$ and $q(x)$ are parameters of the coset manifold $SO(2,d)/SO(1,d)$
which is AdS$_{d+1}$ (this is the so called `solvable subgroup' parametrization
of AdS$_{d+1}$ \cite{solv}). Eq.\p{confih} now yields
\bea
\omega_{\hat{D}} = 0 \quad \Rightarrow \lambda_\mu =
e^{mq}\,\frac{\partial_\mu q}{1 + \sqrt{1 - {1\over 2}e^{2mq}(\partial
q\partial q)}}~, \; \lambda^\mu \equiv
\Lambda^\mu\, \frac{\mbox{tanh}
\sqrt{\frac{\Lambda^2}{2}}}{\sqrt{\frac{\Lambda^2}{2}}}~.\label{adsIH}
\eea
The remaining coset space Cartan forms are then given by the expressions:
\bea \label{adsformsIH}
\omega_P^\mu= e^{-mq}\left( \delta^\mu_\nu -
\frac{\lambda^\mu \lambda_\nu}{1+\frac{\lambda^2}{2}}\right)dx^\nu
\equiv E^\mu_\nu dx^\nu
= e^{-mq}\hat{E}^\mu_\nu dx^\nu\,, \;\omega_K^\mu=\frac{1}{1-\frac{\lambda^2}{2}}
\left( d\lambda^\mu - m \lambda^2 \omega_P^\mu \right).
\eea
The covariant derivative of the Goldstone field $\lambda^\mu$ is
defined by
\bea
\omega^\nu_K=\omega^\mu_P \cD_\mu \lambda^\nu \;\Rightarrow \;
\cD_\mu \lambda^\nu=\frac{1}{1-\frac{\lambda^2}{2}}\left[ e^{mq}\left(\delta_\mu^\rho
+ \frac{\lambda_\mu \lambda^\rho}{1-\frac{\lambda^2}{2}}\right)\partial_\rho\lambda^\nu -
  m\lambda^2 \delta_\mu^\nu \right].\label{covLam}
\eea

The transformation laws of $x^\mu, q(x)$ under the left shifts of
\p{adscoset} are as follows
\bea
\delta x^\mu = a^\mu + c\,x^\mu + 2\,(xb)x^\mu - x^2\,b^\mu +
{1\over 2m^2}\,e^{2mq} b^\mu~,
\;\;\delta q = {1\over m}(c +2\,xb)~. \label{modconf}
\eea
After a field redefinition, they are recognized as the field-dependent conformal
transformations of refs. \cite{mald}, \cite{5}-\cite{nscm} representing
the AdS isometries in the solvable-subgroup parametrization \cite{solv}.

The simplest invariant of the nonlinear realization considered is again
the covariant volume of $x$-space
obtained as an integral of wedge product of $(p+1)$ 1-forms $\omega_P^\mu$.
This invariant is basically
the static-gauge Nambu-Goto (NG) action for $p$-brane in AdS$_{p+2}$
\bea\label{NGaction}
S_{NG} = \int d^{(p+1)} x \left[e^{-(p+1)mq} - \mbox{det}\,E\right]
= \int d^{(p+1)} x\, e^{-(p+1)mq}
\left[1 -\sqrt{1 - {1\over 2}e^{2mq}(\partial q\partial
q)}\right]\,,
\eea
where we have  subtracted 1 to obey the standard requirement of absence
of the vacuum energy \cite{mald}.
The subtracted term is invariant under \p{modconf} on its own
(up to a shift of the integrand by a total derivative).
The action \p{NGaction} is universal, in the sense that it describes the radial (pure AdS) part
of any $(n-2)$-brane action on AdS$_n\times S^m$.

Note that the covariant derivative \p{covLam} which plays an important role in our construction
is the tangent-space projection of the first extrinsic curvature $K_{\mu\nu}$ of the brane
in the static gauge \cite{0} (for the definition of $K_{\mu\nu}$, see e.g.\cite{pol}).
In the flat $m=0$ case
\bea
K_{\mu\nu} = {1\over \sqrt{2}}\frac{1}{\sqrt{1-{1\over 2}(\partial q\partial q)}}\,
\partial_\mu\partial_\nu q\,, \quad
{\cal D}_\mu \lambda_\nu = {1\over \sqrt{2}}(E^{-1})_\mu^\rho (E^{-1})_\nu^\omega
K_{\rho\omega}~.
\eea
The generalization to the AdS case is straightforward.

\section{An equivalence relation between CFT and AdS bases}
In both nonlinear realizations described above we deal with the same coset manifold,
namely $SO(2,d)/SO(1,d-1)$,
in which the coset parameters are separated into the space-time coordinates and Goldstone
fields in two different ways. Hence, there should exist a relation
between these two parametrizations. It can be read off
by comparing \p{confcoset} and \p{adscoset}:
\be
y^\mu=x^\mu-\frac{e^{mq}}{2m}\lambda^\mu \;, \;
\varphi=mq+\ln\left(1-\frac{\lambda^2}{2}\right)\; , \;
\Omega^\mu=m\lambda^\mu \; . \label{map}
\ee
It is invertible at any finite and non-zero $m = 1/R$ and maps the Minkowski space conformal
transformations \p{trstand} onto the field-dependent ones \p{modconf}.
This AdS/CFT transform can be defined only in the framework of extended coset manifolds
$\{y^\mu,\varphi, \Omega^\mu\}$ and $\{x^\mu, q, \lambda^\mu\}$. In Sect. 4 we shall see
how \p{map} can be recovered in the setting with all coset parameters
treated as independent coordinates.

Using \p{map}, any Minkowski space conformal field theory
with a dilaton among its basic fields can be projected onto the variables
of AdS brane and vice versa. To find the precise form of various $SO(2,d)$
invariants in both bases, let us define the transition matrix
\bea
\frac{\partial y^\nu}{\partial x^\mu}\equiv
  {\cal A}_\mu^\nu=\delta_\mu^\nu -\frac{\lambda_\mu \lambda^\nu}{1+\frac{\lambda^2}{2}} -
   \frac{e^{mq}}{2m}\partial_\mu \lambda^\nu =
\left(1-\frac{\lambda^2}{2}\right)\hat{E}_\mu^\rho\, T^\nu_\rho\;,\;
T^\nu_\rho = \delta^\nu_\rho - {1\over 2m}{\cal D}_\rho\lambda^\nu~.\label{defT}
\eea
The Jacobian of the change of space-time coordinates in \p{map} is
\be
J \equiv \mbox{det}\,{\cal A} = \left(1-\frac{\lambda^2}{2}\right)^{p+1}\mbox{det}\,\hat{E}\,
\mbox{det}\,T~. \label{Jac}
\ee
Then, making the change of variables \p{map} in the invariant
dilaton Lagrangians \p{S1} and \p{Skin1},
we obtain, respectively,
\bea
S_1 &=&
\int d^{(p+1)}x\,e^{-(p+1)m q}\,\sqrt{1 -{1\over 2}e^{2mq}(\partial q \partial q)}\;
\mbox{det}\, T~,
\label{S1transf} \\
S_\varphi^{kin} &=&
m\,\int d^{(p+1)}x\,e^{-m(p+1)q}\,\sqrt{1 -{1\over 2}e^{2mq}(\partial q \partial q)}\,
\left[\mbox{det} T\,(T^{-1}{\cal D}\lambda)^\mu_\mu \right].
\label{Skintransf}
\eea
A surprising fact is that the AdS image of the potential term of dilaton contains
the NG part of the AdS $p$-brane action \p{NGaction} modified by higher-derivative covariants
collected in $\mbox{det} (I - {1\over 2m}{\cal D}\lambda) =
1 -{1\over 2m}{\cal D}_\mu\lambda^\mu +\ldots\;$.
As we saw, ${\cal D}_\mu\lambda^\nu$ is basically the $p$-brane extrinsic curvature.
So, already the simplest conformal invariant in Minkowski space proves to produce, on
the AdS side,
an action which is the standard $p$-brane action in AdS$_{p+2}$ plus corrections composed
of the extrinsic curvature tensor. Note that only for the conformal actions containing
no potential terms of dilaton the relations \p{map} can be treated as a genuine equivalence map
taking the kinetic term of $\varphi$ into that of $q$ (plus higher order corrections).

Let us now see what the brane action \p{NGaction} looks like in the conformal basis. Using
\be
{\cal D}_\mu\Omega^\nu =m(T^{-1})_\mu^\omega {\cal D}_\omega\lambda^\nu~,
\quad (T^{-1})_\mu^\nu = \delta^\nu_\mu + {1\over 2m^2}{\cal
D}_\mu\Omega^\nu~.
\ee
and making in \p{NGaction} the change of variables inverse to \p{map}, we find
\be
S_{NG} = {1\over 4m^2}\int d^{(p+1)}y\,e^{(1-p)\varphi}\,
\frac{(\partial \varphi\partial\varphi)}{1-{1\over 8m^2}e^{2\varphi}
(\partial \varphi\partial \varphi)}\,
\mbox{det}\left(I + {1\over 2 m^2}{\cal D}\Omega \right).\label{NGconf}
\ee
Thus we have found an equivalent representation of the static-gauge action \p{NGaction}
of $p$-brane in AdS$_{p+2}$ as a non-linear extension of the conformally-invariant
dilaton action in ${\cal M}_{p+1}$. In \cite{0} the conformal field theory image
of the full bosonic part of
D3-brane action on AdS$_5\times S^5$ was found.

\section{Geometric interpretation of the AdS/CFT map in a conformal space with extra dimensions}
As was already mentioned in the previous Section, the transformation \p{map} cannot be understood
within the pure $\{y, \varphi \}$, or $\{x, q\}$ geometries. Indeed, the extended manifold
$\{y^\mu, \varphi(y)\}$ has the topology of ${\cal M}_d \times R^1$, while $\{x^\mu, q(x) \}$
is a surface in AdS$_{d+1}$. Clearly, no direct equivalence can be established between
these two geometrically different manifolds.

The meaning of \p{map} can be clarified by embedding both these manifolds as subspaces into
the extended conformal space $CM^{2d+1} \equiv \{y^\mu, \varphi, \Omega^\nu \} =
\{x^\mu, q, \lambda^\nu \}$.
This is the coset $SO(2,d)/SO(1,d-1)$ in which all parameters are treated as {\it independent}
coordinates. The nonlinear realizations associated with the parametrizations
\p{confcoset} and \p{adscoset}
operate on $d$-dimensional hypersurfaces in this ambient space, parametrized, respectively,
by $y^\mu$ and $x^\mu$.

The $SO(2,d)$ transformation properties of the coordinates of $CM^{2d+1}$
in the conformal and AdS parametrizations can be obtained as before by considering left action
of $SO(2,d)$ on the coset elements \p{confcoset} and \p{adscoset}, in which all parameters
are independent.
The coset Cartan 1-forms in eq. \p{confcartan} and their counterparts in the AdS basis define
covariant
differentials of the coordinates $\{y^\mu, \varphi, \Omega^\nu \}$ and
$\{x^\mu, q, \lambda^\nu \}$,
respectively. The Lorentz Cartan form in \p{confcartan} defines the $SO(1,d-1)$ connection
which enters
the covariant differentials of those functions on $CM^{2d+1}$ which have external
indices with respect to the stability sibgroup $SO(1,d-1)$ and so carry non-trivial
representations of the latter.

To clarify what is the meaning of the transformation \p{map} in this setting, let us consider
a scalar function on $CM^{2d+1}$, first in the conformal parametrization,
\be
F(y, \varphi, \Omega)\,, \label{Func}
\ee
and study the issue of existence of invariant subspaces in $CM^{2d+1}$. This problem amounts
to listing all possible covariant conditions which one can impose on $F$ to effectively suppress
the dependence of $F$ on one or another coordinate of $CM^{2d+1}$. The covariant derivatives
of the function $F$ are defined by the standard formula
\be
d F(y,\varphi, \Omega) = dy^\mu\,\partial^y_\mu F + d\Omega^\nu\,\partial^\Omega_\mu F
+ d\varphi\,\partial^\varphi F
\equiv \omega^\mu_P\,\nabla_\mu^y F + \omega^\mu_K\,\nabla_\mu^\Omega F
+ \omega_D\,\nabla^\varphi F\,, \label{defCovD}
\ee
whence
\bea
\nabla^y_\mu = e^\varphi \partial/\partial y^\mu +
2\Omega_\mu\,\partial/\partial \varphi + \Omega^2\,\partial/\partial \Omega^\mu\,,\;
\nabla^\varphi = \partial/\partial \varphi + \Omega^\mu\partial/\partial \Omega^\mu\,, \;
\nabla^\Omega_\mu = \partial/\partial \Omega^\mu\,. \label{CovD}
\eea
While acting on a function with an external Lorentz $SO(1,d-1)$ index,
the covariant derivative $\nabla^y_\mu$ acquires a
Lorentz connection which is determined by the Cartan form associated with the generators
$M_{\mu\nu}$ in \p{confcartan}.
E.g., the covariant derivative of a vector function $G_\mu $ is defined as
\be
{\cal D}^y_\mu \,G_\nu = \nabla^y_\mu\,G_\nu +
2 (\Omega_\nu\, G_\mu - \eta_{\mu\nu} \Omega^\rho G_\rho)\,.
\ee
Other derivatives do not acquire any connections.

The evident chain of the subspaces closed under the action of $SO(2,d)$
\be
\{y, \varphi, \Omega\} \;\;\supset\;\; \{y, \varphi \} \;\;\supset\;\; \{y \} \label{chain}
\ee
is in the one-to-one correspondence with the following $SO(2,d)$ covariant analyticity-type
constraints on the generic function $F(y, \varphi, \Omega)$:
\bea
(a)\; \nabla^\Omega_\mu F^{(1)} = 0 \; \Rightarrow F^{(1)} = f(y, \varphi)\,,\;\;
(b)\; \nabla^\Omega_\mu F^{(2)} = \nabla^\varphi F^{(2)} = 0
\;\Rightarrow F^{(2)} = f(y)\,.\label{chain2}
\eea
The self-consistency of these constraints follows from the integrability conditions
\be
\nabla^\Omega_\mu \nabla^\Omega_\nu - (\mu \leftrightarrow \nu) = 0\,,
\quad  [\nabla^\varphi, \nabla^\Omega_\mu] = -\nabla^\Omega_\mu\;.
\label{compat}
\ee
Note that the field $f(y)$ in \p{chain2} has the conformal weight zero. In order to end up with
a scalar field having the standard free field weight $(d-2)/2$, one should replace the second
set of constraints in \p{chain2} by
\be
\nabla^\Omega_\mu \, F^{(2)} = 0\,, \;\; \nabla^\varphi\,F^{(2)} = {1\over 2}(d-2)\,F^{(2)}
\quad \Rightarrow
\quad F^{(2)} = e^{({d\over 2} -1)\varphi}\,\tilde{f}(y) \label{modif}
\ee
($e^\varphi$ has the weight $-1$). This choice of covariant constraints
is also self-consistent thanks to \p{compat}.
In the considered parametrization all these constraints are easily solved just because
the corresponding
covariant derivatives are basically partial derivatives with respect to the appropriate coordinates
and \p{chain2}, \p{modif} simply eliminate (or strictly fix) the dependence on the latter.

An important outcome of the above discussion is that the conformal parametrization
$\{y, \varphi, \Omega \}$
manifests the embedding chain \p{chain} which corresponds to splitting of $CM^{2d+1}$
into the product of
the base Minkowski space ${\cal M}_d = \{y\}$ and the fiber $\{\varphi, \Omega\}$,
with $SO(2,d)$ being realized in ${\cal M}_d$ as the standard conformal group. Let us now
impose on the generic function $F$ a different type of the covariant constraint
\be
\left(\nabla^\Omega_\mu + \alpha \nabla^y_\mu \right)F^{(3)} =0\,, \label{mixed}
\ee
where $\alpha $ is a constant. This constraint is again self-consistent due to
the integrability conditions
\be
[\, \nabla^\Omega + \alpha \nabla^y,  \nabla^\Omega + \alpha \nabla^y\, ]
\sim  \nabla^\Omega + \alpha \nabla^y\,,
\ee
or, equivalently,
$$
(\nabla^\Omega_\mu + \alpha {\cal D}^y_\mu)(\nabla^\Omega_\nu +\alpha  \nabla^y_\nu)
- (\mu \leftrightarrow \nu) = 0\,.
$$
At $\alpha = 0$ and $\alpha = \infty$ this mixed constraint goes over to its counterpart from the
set \p{chain2} and another admissible constraint related to \p{chain2} by
conformal inversion. Note that at $\alpha \neq 0, \infty$ one cannot impose on $F$ any
additional constraint involving $\nabla^\varphi\,$, since
\be
[\nabla^\varphi, \nabla^\Omega_\mu + \alpha \nabla^y_\mu ] \sim  \nabla^\Omega_\mu
- \alpha \nabla^y_\mu\,,
\ee
and so $\nabla^\varphi $ does not constitute a closed subalgebra with the
differential operator in \p{mixed}.
To understand the meaning of \p{mixed} one should pass to such a parametrization
of $CM^{2d+1}$ in which
the differential operator in \p{mixed} is reduced to a partial derivative, suggesting
that $F$ subjected to \p{mixed} is independent of the relevant coordinate. Identifying
\be
\alpha = - 1/2m^2
\ee
and performing the coordinate change just according to \p{map},
it is straightforward to find how the
covariant derivatives look in the new coordinates
$\{x^\mu, q, \lambda^\mu \}$. In particular, we find
\be
\nabla^\Omega_\mu  - {1\over 2m^2}\nabla^y_\mu =
{1\over m}\left(1 -\frac{\lambda^2}{2} \right)
\frac{\partial}{\partial \lambda^\mu}\,.\label{CRads}
\ee
Thus the basis $\{x, q, \lambda \}$ is just the one in which the differential constraint
\p{mixed} takes the
`short'  analyticity condition type form and so becomes explictly solvable:
\be
F^{(3)}(y, \varphi, \Omega) \equiv \tilde{F}^{(3)}(x, q, \lambda) = \tilde{f}(x, q)\,. \label{F3}
\ee
We know that $x^\mu $ and $q$ provide a parametrization of AdS$_{d+1}$,
so the basis $\{x, q, \lambda \}$
in $CM^{2d+1}$ makes manifest the embedding
\be
CM^{2d+1} \;\;\supset \;\; \mbox{AdS}_{(d+1)}\,,\label{adschain}
\ee
where the subspace AdS$_{d+1}$ is again closed under the action of $SO(2,d)$ which is realized by
the transformations \p{modconf}. As distinct from the chain \p{chain}, one cannot extract
any subspace in AdS$_{d+1}$ which would be closed under $SO(2,d)$. In the conformal basis
this property is rephrased as the impossibility to strengthen \p{mixed} by any additional
constraint with $\nabla^\varphi$.

From the mathematical point of view, the embedding chains \p{chain} and \p{adschain}
(as well as some other possible ones \footnote{One more interesting invariant subspace
of $CM^{2d+1}$ is the $2d$-dimensional `bi-conformal space' \cite{niedW} which is obtained by
placing the generator $D$ into the stability
subgroup. It corresponds to imposing the single constraint $\nabla^\varphi F =0$.})
amount to different fiberings of the coset
manifold $CM^{2d+1} = SO(2,d)/SO(1,d-1)$. The option \p{chain}
corresponds to the choice of ${\cal M}_d = \{y^\mu \}$ as the base and $\{q, \Omega^\mu\}$
as a fiber, while in the case \p{adschain} the base and fiber are AdS$_{d+1}$ and
the coset $SO(1,d)/SO(1,d-1) = \{\lambda^\mu \}$, respectively. Also notice that
one could recover the variable change \p{map}, up to an equivalence transformation
$q \rightarrow \tilde{q}(q)$, $\lambda^\mu \rightarrow \tilde{\lambda}^\mu(\lambda, q)$,
simply by requiring the covariant derivative in \p{CRads} to have the `short' form
$$
\nabla^\Omega_\mu  - {1\over 2m^2}\nabla^y_\mu =
A_\mu^\nu (q, \lambda)\, \frac{\partial}{\partial \lambda^\nu}\,,\;\;
A_\mu^\nu = \delta^\nu_\mu + O(q,\lambda)\,,\;\; \mbox{det}\, A|_{q=\lambda =0} \neq 0\,.
$$

To summarize, the change of coordinates \p{map} defines passing from the parametrization
$\{y, \varphi, \Omega \}$
of $CM^{2d+1}$ in which the Minkowski space geometry is manifest and $SO(2,d)$ acts as
the conformal group of ${\cal M}_d$ \footnote{To be more rigorous, of the appropriate
compactification of ${\cal M}_d$.} to the parametrization $\{x, q, \lambda \}$ where
the AdS$_{d+1}$ geometry is manifest and $SO(2, d)$ is realized as the corresponding group
of motion. In the original setting this transformation relates two different
$d$-dimensional hypersurfaces in $CM^{2d+1}$ which are parametrized, respectively, by
the coordinates $y^\mu$ and $x^\mu$. The first hypersurface is pertinent to
the standard nonlinear
realization of $SO(2,d)$ on $y^\mu$ and dilaton field $\varphi(y)$, while the second one is
just the worldsurface of scalar $(d-1)$-brane on AdS$_{d+1}$ in a static gauge.

Besides relating Minkowski space conformal theories in spontaneously broken phase to AdS branes,
the transformation \p{map}
seems to have some interesting potential implications for the field theories on AdS spaces.
This follows from
the consideration of the present Section. Indeed, using \p{map} one can represent
any unconstrained
field on AdS$_{d+1}$ in the solvable subgroup parametrization, $f(x,q)$, as a constrained
field on $CM^{2d+1}$, i.e. $F^{(3)}(y,\varphi, \Omega)$ subjected to the constraint \p{mixed}
(with $\alpha = -1/ 2m^2$).\footnote{This resembles e.g. the description
of chiral superfields in $N=1, d=4$ supersymmetry either as unconstrained functions on the chiral
$N=1$ superspace or constrained functions on real $N=1$ superspace. Both these superspaces are
subspaces of general complexified $N=1, d=4$ superspace which palys a role similar
to $CM^{2d+1}$ in our
case.}Assuming for $F(y,\varphi, \Omega)$ a series expansion in $\Omega^\mu$,
$F(y,\varphi, \Omega) = F^0(y, \varphi) + F_\mu(y, \varphi)\Omega^\mu +
{1\over 2}F_{(\mu\nu)}(y,\varphi)
\Omega^\mu\Omega^\nu + O(\Omega^3)$, eq. \p{mixed} expresses the whole infinite tower
of symmetric tensor
coefficients in such an expansion as multiple $y$- and $\varphi$-derivatives of $F^0(y, \varphi)$:
$$
F_\mu = {1\over 2 m^2}\,e^\varphi\,\partial_\mu\,F^0\,, \quad F_{\mu\nu} =
{1\over 4 m^4}\,e^{2\varphi}\,
\partial_\mu\partial_\nu F^0 + {1\over m^2} \eta_{\mu\nu}e^\varphi \partial_\varphi\,F^0\,,
$$
etc. Thus  the AdS$_{d+1}$ field $f(x,q)$ proves to be equivalent to an infinite set of
conformal fields on the Minkowski space ${\cal M}_d = \{y^\mu \}$ emerging from
the appropriate expansion of
$F_0(y,\varphi)$ in the dilaton-like coordiante $\varphi$. Assuming that the correct
expansion should
be a general series in the positive and negative powers of $z\equiv e^\varphi$ with
the conformal dimension
$-1$, we conclude that these fields should carry all integer conformal dimensions
from $-\infty$ to $\infty $ (some truncations are possible depending
on the boundary conditions with respect to $z$ or $q$).
Conversely, some irreducible conformal field on ${\cal M}_d$ can be represented
as a constrained field on $CM^{2d+1}$ in the AdS parametrization, and so it amounts to a set of
fields on AdS$_{d+1}$ with the properly restricted dependence on $\{x^\mu, q \}$.
It is worth noting that all these relationships are purely `kinematical'.

As for dynamics, the existence of the map \p{map} offers an interesting new opportunity in analysing
the relationships between equations of motion for fields on AdS$_{d+1}$ and conformally invariant
equations in ${\cal M}_d$, including those for higher spin fields. The covariant dynamical equations
for fields of any spin on AdS were recently constructed in \cite{Misha1}.
One more link with higher spins is suggested by the fact that general functions
on $CM^{2d+1}$ contain in their $\Omega^\mu$- or $\lambda^\mu$-expansions infinite sequences
of symmetric Lorentz
tensor fields  which are basic ingredients of various versions of the higher integer spin theories
(see e.g. review \cite{Misha2} and \cite{Segal}). It would be interesting to study whether
these theories admit a reformulation in $CM^{2d+1}$ and what are possible implications
of the transformation \p{map} in this context.

The above reasoning shows some important difference in the treatment of the relationship
between $d$-dimensional Minkowski and AdS$_{d+1}$ spaces in the conventional AdS/CFT approach
and in the
considered framework. While in the first approach the (compactified) ${\cal M}_d$ is regarded
as a boundary of AdS$_{d+1}$, in the case under consideration these both manifolds coexist
as different subspaces in the extended conformal
space $CM^{2d+1}$. The coordinate map \p{map} simply relates two different parametrizations
of $CM^{2d+1}$ which make manifest
either ${\cal M}_d$ geometry or AdS$_{d+1}$ geometry. Note that one can equally relate
the conformal parametrization
of $CM^{2d+1}$ to the parametrization which manifests the geometry associated with the subspace
$SO(2,d)/SO(2,d-1) \subset CM^{2d+1}$. The corresponding invariant functions
are singled out by the constraint \p{mixed} with $\alpha \rightarrow -\alpha$. Obviously,
there is also
a change of coordinates from this new parametrization to the AdS one. The subspace just mentioned
has as its flat limit the `two-time' $(d+1)$-dimensional space
(with the signature $(++, - -\ldots)$),
so some interrelations with the `two-time' physics \cite{bars} are expected to arise
while exploring these maps and their consequences. One of such consequences is the possibility
to relate AdS branes to those on this exotic manifold, and vice versa.

Finally, let us notice that the covariant derivatives applied to the functions of the type
\p{chain2}, \p{modif} or \p{F3} in general take them out of the subspaces on which they
are defined.
Only those covariant operators which {\it commute} with the analyticity conditions
\p{chain2}, \p{mixed}
preserve the type of a given constrained function. A technically feasible way to construct such
operators is to exploit invariance with respect to the appropriate right transformations
of the coset
parameters. An equivalent way to covariantly restrict general functions on $CM^{2d+1}$ to
the invariant subspaces is to require these functions to be invariant under right
shifts of the coset elements \p{confcoset} or \p{adscoset} by the generators which enlarge
the stability subgroup
$SO(1,d)$ to the stability subgroups of these subspaces viewed as coset manifolds.
The differential
operators appearing in the constraints \p{chain2}, \p{modif}, \p{F3} prove to be generators
of these
right shifts, and the constraints themselves admit a nice interpretation as the conditions
of invariance under these shifts (or as a condition that the given field is an eigenfunction of
some Cartan generator of the group of right shifts, as in \p{modif}). Then the precise form of
covariant differential operators preserving the given type of constrained function can also be
found from the requirement of invariance with respect to the right shifts.

For instance, the first constraints in \p{chain2}, \p{modif}  amount to invariance under
the right transformations
with the generator $K^\mu $. Using \p{confcartan}, it is very easy to find how
these right transformations
are realized on Cartan forms, coset coordinates and covariant derivatives:
\be
\delta \Omega^\mu = \beta^\mu\,, \;\delta \nabla^\Omega_\mu = 0\,, \; \delta \nabla^\varphi =
\beta^\mu\,\nabla^\Omega_\mu, \; \delta \nabla^y_\mu = 2 \beta_\mu \nabla^\varphi +
2[\,(\Omega\cdot \beta)\delta^\nu_\mu -
\beta_\mu\Omega^\nu\,]\,\nabla^\Omega_\nu\,, \label{Rtr}
\ee
where $\beta^\mu$ is the corresponding group parameter. Then it is easy to check that
the covariant d'Alembertian $\Box^{(d)}_{cov} \equiv {\cal D}^{y\,\mu}\nabla^y_\mu$,
being applied to
the functions subjected to the constraints \p{modif}, is invariant under these
transformations and so cannot
depend on $\Omega^\mu$ (while such a dependence is present for generic functions
on $CM^{2d+1}$). One finds
\be
\Box_{cov}^{(d)}\,F^{(2)} \equiv  {\cal D}^{y\,\mu}\nabla^y_\mu \,F^{(2)} =
e^{({d\over 2} -1)\varphi}\,\Box^{(d)}
\tilde{f}(y)\,. \label{Dal1}
\ee

The AdS$_{d+1}$ case \p{mixed} is more complicated because the extra right transformations
in this case are generated by $\hat{K}^\mu = m K^\mu - {1\over 2m}P^\mu$
which enlarges $SO(1,d-1)$ to the non-ableian stability subgroup $SO(1, d)$ of AdS$_{d+1}$.
Nevertheless the corresponding analogs of \p{Rtr} can be found in this case too,
and an analog of the covariant d'Alembertian \p{Dal1} can be uniquely determined from
the condition of invariance under these transformations. It is constructed from
the covariant derivatives as follows (with taking account of \p{mixed})
\be
\Box_{cov}^{(d+1)} = {1\over 2}\left({\cal D}^{y\,\mu} +
2m^2\,\nabla^{\Omega\,\mu}\right)\nabla^y_\mu -
2m^2\,\nabla^\varphi \nabla^\varphi
\ee
and in the basis $\{x, q, \lambda^\mu \}$ it is independent of $\lambda^\mu$  when acts on
the functions subjected to \p{mixed}
\be
\Box_{cov}^{(d+1)}\,F^{(3)}(y,\varphi, \Omega) = \left( e^{2mq} \Box^{(d)}
- 2 \frac{\partial^2}{\partial q^2} + 2md\;
\frac{\partial}{\partial q}\right) f(x,q)\,.
\ee
It is just the covariant d'Alembertian of a scalar field on AdS$_{d+1}$
in the considered parametrization. It
is sraightforward to check its invariance under the transformations \p{modconf}.

\section{The d=1 case: (super)conformal mechanics revisited}
Conformal mechanics (CM) \cite{dff} and its superconformal extensions \cite{AP} are
the simplest models of (super)conformal field theory.
Recently, it was suggested \cite{nscm} that the so-called `relativistic'
generalizations
of these $d=1$ models are candidates for the conformal field theory
dual to AdS$_2$ (super)gravity in the AdS$_2$/CFT$_{1}$ framework.
The simplest model of this kind is a charged particle evolving on
the AdS$_2\times S^2$
background (the Bertotti-Robinson metric). It describes a near-horizon geometry of
the extreme $d=4$ Reissner-Nordstr\"om black hole. The action
(or Hamiltonian) of the standard CM
can be recovered from the worldline action (or Hamiltonian) of the `relativistic' CM
model in the `weak-field' (or `small velocity') approximation.

Both the `old' and `new' (super)conformal mechanics models respect
the same (super)\break conformal symmetry,
which suggests that these models can in fact be equivalent to each other.
The $d=1$ version of the
equivalence map \p{map} allows one to explicitly prove this conjecture.

The `old' CM can be described in terms of nonlinear realization of the $d=1$ conformal group
$SO(2,1)$ \cite{we}. The $so(2,1)$ algebra is
\be\label{confbasis1}
\left[ P,D\right] =-P\; , \; \left[ K,D\right] = K\; , \; \left[ P,K \right] =-2D\; .
\ee
One defines a nonlinear realization of $SO(1,2)$ as left shifts of the element
\be\label{gconf}
g=e^{tP} e^{u(t)D} e^{\lambda(t) K}~.
\ee
The $SO(2,1)$ left shifts induce for $t, u(t)$ and $\lambda(t)$ the following
transformations
\be
\delta t = a +b\,t + c\,t^2 \equiv a(t) ~, \quad
\delta u = \dot a(t) = b + 2c\,t ~, \quad
\delta \lambda = c\,e^u ~. \label{nel1tran}
\ee
The left-invariant Cartan forms are defined by
\bea\label{formsconf}
g^{-1}d g = \omega_P P +\omega_D D + \omega_K K =
e^{-u}dt P + (du-2 e^{-u} \lambda dt)D
+(d\lambda + e^{-u}\lambda^2 dt -\lambda du)K\,.
\eea
The coset field $\lambda(t)$ can be covariantly eliminated by the constraint
\be\label{IHconf}
\omega_D=0 \; \Rightarrow \; \lambda =\frac{1}{2}e^u \dot{u}
\;.
\ee
Then the manifestly invariant worldline action
\bea
S = -{1\over 2} \int \left(\mu \omega_k + \gamma\, \omega_P\right)
= {1\over 2} \int dt\left({1\over 4}\,\mu\,e^u\, {\dot{u}}^2
-\gamma\, e^{-u} \right)~, \label{nel1act}
\eea
upon the identificationis $x(t) = e^{1/2\, u(t)}$ is just the `old'
conformal mechanics action \cite{dff},
\be
S = {1\over 2} \int dt \left( \mu \,\dot{x}^2 - {\gamma\over x^2} \right)~. \label{cmact}
\ee

Let us now pass to the AdS$_2$ basis in \p{confbasis1}, redefining $K$ and $D$ as
\be\label{adsgenerators2}
\hK =mK -\frac{1}{m}P\;,\; \hD=mD \;.
\ee
An element of $SO(2,1)$ in the AdS basis is defined as
\be\label{confcoset2}
g=e^{yP}e^{\phi(y) \hD} e^{\Omega(y) \hK} \; .
\ee
The parameters $y, \phi$ represent AdS$_2 \sim SO(2,1)/SO(1,1)$ in the solvable
subgroup parametrization:
\be
\delta y = a(y) + \frac{1}{m{}^2}\,c\,e^{2m\phi}~, \quad \delta \phi = \frac{1}{m}\,
\dot{a} = \frac{1}{m}\,(b+2c\,y)~, \quad \delta \Omega = \frac{1}{m}\,c\,e^{m\phi} ~.
\label{modphitr}
\ee
The Cartan forms in the AdS parametrization are related to \p{formsconf} as
\be
\omega_K = m\,\hat\omega_K~, \quad \omega_P = \hat\omega_P - {1\over
m}\,\hat\omega_K~, \quad \omega_D = m\,\hat\omega_D~. \label{rel222}
\ee
Like $\lambda(t)$ in eq. \p{IHconf}, the field $\Lambda(y) = \tanh \Omega(y)$
can be covariantly eliminated
\bea
&& \hat\omega_D=0 \; \Rightarrow \; \Lambda = \dot\phi\,e^{m\phi}\,\frac{1}{1 + \sqrt{1 -
\dot\phi^2\,e^{2m\phi}}}~,\;\;\hat\omega_P = e^{-m\phi}\,\sqrt{1 - e^{2m\phi}\,\dot\phi^2 }\;dy~,
\label{Lamb} \\
&& \hat\omega_K = -\frac{m}{2}\,e^{-m\phi}\left(1 - \sqrt{1 - e^{2m\phi}\,\dot\phi^2
}\right)dy
+ \mbox{Total derivative}\times dy~. \label{invHforms}
\eea
The invariant action for $\phi(y)$ can now be easily constructed by substituting
the expressions \p{invHforms} for $\omega_P, \omega_K$ in \p{nel1act} using the relation \p{rel222}:
\be
S = \int \left[ (q-\tilde\mu)\,\hat\omega_P - (2/m)q\,\hat\omega_K \right] =
\int dy\, e^{-m\phi}\left(q - \tilde\mu\, \sqrt{1 -  e^{2m\phi}\,\dot\phi^2}\right), \label{invact2}
\ee
where
\be
q= {1\over 4} (m^2 \mu - \gamma)~, \quad \tilde{\mu} = {1\over 4}(m^2 \mu
+ \gamma)~. \label{identif1}
\ee
After a field redefinition, \p{invact2} is recognized as the radial-motion part of the `new'
CM action of ref. \cite{nscm}. The same result could be
equivalently obtained by performing in \p{nel1act} the $d=1$ AdS/CFT
transformation obtained by comparison of \p{gconf} and \p{confcoset2}.
\be
t=y-\frac{1}{m}\,e^{m\phi}\Lambda\;,\; u=m\,\phi +\ln (1-\Lambda^2)\; ,\;
\lambda=m\,\Lambda \;. \label{conn}
\ee

Let us briefly discuss (basically following \cite{00}) how this correspondence can be generalized
to SCM models. We consider $N=2$ SCM as the simplest case.

The starting point is the $su(1,1|1)$ superalgebra which includes, apart from the $so(1,2)$ generators
\p{confbasis}, those of Poincar\'{e} $\left\{ Q,\bQ \right\}$ and conformal
$\left\{ S,\bS \right\}$ supersymmetries and the $U(1)$ generator $U$. In the
conformal basis the non-vanishing (anti)commutators read:
\bea\label{superalgconf}
&& \left\{ Q,\bQ\right\}=2i P\;,\;\left\{ Q,\bS\right\}=2iD -2i U\;,\;
\left\{ S,\bS\right\}=2iK\;,\;\left\{ S,\bQ\right\}=2iD +2iU\;,\nn
&& \left[ P, \left( \begin{array}{c} S \\ \bS \end{array} \right) \right]= -
   \left( \begin{array}{c} Q \\ \bQ \end{array} \right) \; , \;
   \left[ K, \left( \begin{array}{c} Q \\ \bQ \end{array} \right) \right]=
   \left( \begin{array}{c} S \\ \bS \end{array} \right) \; , \nn
&& \left[ D, \left( \begin{array}{c} Q \\ \bQ \end{array} \right) \right]= \frac{1}{2}
   \left( \begin{array}{c} Q \\ \bQ \end{array} \right) \; , \;
   \left[ D, \left( \begin{array}{c} S \\ \bS \end{array} \right) \right]= -\frac{1}{2}
   \left( \begin{array}{c} S \\ \bS \end{array} \right) \; , \nn
&& \left[ U, \left( \begin{array}{c} Q \\ \bQ \end{array} \right) \right]= \frac{1}{2}
   \left( \begin{array}{r} Q \\ -\bQ \end{array} \right) \; , \;
   \left[ U, \left( \begin{array}{c} S \\ \bS \end{array} \right) \right]= \frac{1}{2}
   \left( \begin{array}{r} S \\ -\bS \end{array} \right) \; .
\eea

The standard nonlinear realization of $SU(1,1|1)$ as the $d=1, N=2$ superconformal group
is set up as left multiplications of the coset
\be\label{superGconf}
g=e^{tP}e^{\theta Q+\bt\bQ}e^{qD}e^{\lambda K}e^{\psi S+ \bpsi\bS} \;,
\ee
where $(t, \theta, \bar\theta)\equiv z$ are coordinates of $d=1, N=2$ superspace and
the remaining coset parameters are superfields given on this superspace. The transformation
rules of the supercoset parameters and the structure of the related left-covariant Cartan
superforms can be found in \cite{00}. We
only notice that on the $d=1, N=2$ superspace coordinates one recovers the standard
$N=2$ superconformal transformations, while all the superfield coset parameters
are expressed through the only essential one $q(z)$ by the appropriate inverse Higgs
constraints:
\bea
&&\lambda = \frac{1}{2}\, e^q\, \dot{q}\;, \;\;\bpsi =
-\frac{i}{2}\,e^{\frac{1}{2}q}\,D q\;,\;\;
\psi = -\frac{i}{2}\,e^{\frac{1}{2}q}\,\bD q\;,\label{susyinv1} \\
&& D= \frac{\partial}{\partial\theta}+i\bt \partial_t\;,\; \bD=
\frac{\partial}{\partial\bt}+i\theta \partial_t\;,\; \left\{ D,\bD\right\}
=2i \partial_t \;. \nonumber
\eea
The invariant action of $N=2$ SCM reads
\be\label{superactionconf}
S_{N=2}=\int dt d^2\theta \left[ \frac{\mu}{2}D Y \bD Y +
 \sqrt{\mu\gamma}\; \ln(Y)\right] \;, \quad Y = e^{\frac{1}{2}\,q}\,.
\ee
Its bosonic core coincides with \p{nel1act} upon identification $q|_{\theta = 0} = u$ and
eliminating the auxiliary field $[D,\bD]q|_{\theta = 0}$ by its equation of motion.

Now we shall consider a supersymmetric extension of the AdS basis
\p{adsgenerators}. The only new thing we have to do is to make
a rescaling of the superconformal generators as $\hS=mS, \hbS=m\bS$.
We define the realization of $SU(1,1|1)$ in the AdS basis by its left
action on the coset $SU(1,1|1)/U(1)$ in the following parameterization:
\be\label{superGads}
g=e^{yP}e^{\theta Q+\bt\bQ}e^{\Phi\hD}e^{\Omega \hK}e^{\xi \hS+ \bxi\hbS} \;.
\ee
Like in the case of standard nonlinear realization, one can directly find the transformation
rules of the superspace coordinates and Goldstone superfields. As distinct from the standard case,
the transformation laws of coordinates now essentially include Goldstone superfields, i.e.
we deal with a field-dependent realization of $N=2$ superconformal group which is
a generalization
of the bosonic realization \p{modphitr}. The only essential Goldstone superfield is $\Phi$,
the remaining ones are expressed through $\Phi$ by the corresponding inverse Higgs constraints:
\be
\Lambda = e^{m\Phi}\, \partial_y\Phi\,\frac{1}{1+\sqrt{1-e^{2m\Phi}(\partial_y \Phi)^2}}\;,\;
\;\xi = -\frac{i}{2}\, \frac{1 + \Lambda^2}{\sqrt{1-\Lambda^2}}\,
e^{\frac{m}{2}\Phi}\, \bD_y\Phi\;.
\label{IHexpr2}
\ee

By comparing two different parametrizations of the same coset $SU(1,1|1)/U(1)$,
eqs. \p{superGconf} and \p{superGads}, one can find $N=2$ extension of the
transformation \p{conn}
\be\label{connsusy}
t=y-\frac{1}{m}e^{m\Phi}\Lambda\;,\; q=m\Phi +ln(1-\Lambda^2)\; ,\; \lambda=m\Lambda \;,\;
\psi=m\xi\;,\; \bpsi=m\bxi\;.
\ee

Now we can obtain the invariant superfield action
which is pertinent to the above AdS realization of $d=1, N=2$ superconformal group and so
is expected to describe $N=2$ superextension of the bosonic BR particle action \p{invact2}.
One should perform the transformation \p{connsusy} in the `old' $N=2$ SCM action
\p{superactionconf}. For simplicity, we choose $\gamma = 0$, which amounts
to requiring zero vacuum energy. We obtain
\bea\label{adsaction0}
S= \frac{1}{2} \int dt d^2\theta \left( -\mu \bpsi\psi \right)
 = \frac{\mu m^2}{8} \int dy d^2\theta\,e^{m\Phi}
\left( \frac{1-\Lambda^2}{1+\Lambda^2} -\frac{1}{m}e^{m\Phi}\partial_y \Lambda\right)
\frac{ (1+\Lambda^2)^2}{1-\Lambda^2}
 D_y \Phi \bD_y \Phi
\eea
where $\Lambda$ is expressed through $\Phi$ according to \p{IHexpr2}.

It is straightforward to pass to the component fields in \p{adsaction0} and to show
that, when all fermions are discarded,
\be
F= 0
\ee
on shell. After substituting this into the pure bosonic part of the component action,
$S_{bos}\,$, the latter, modulo a total derivative in the Lagrangian, becomes
\be
S_{bos} = \frac{\mu m^2}{4}\,\int dy e^{-m\phi}\left(1
-\sqrt{1-e^{2m\phi}(\partial_y \phi)^2}\right), \label{bosonA}
\ee
which coincides with \p{invact2} upon the identification \p{identif1} (for $\gamma = 0$).

Thus \p{adsaction0} provides a manifestly $N=2$ supersymmetric off-shell form of
$N=2$ superconformal extension of the `new' conformal mechanics action \p{invact2}
which describes the radial (AdS$_2$) motion of the charged particle in the BR
AdS$_2\times S^2$ background. By construction, it is related by the equivalence transformation
\p{connsusy} to the $\gamma = 0$ case of the `old' $N=2$ superconformal mechanics
action \p{superactionconf}.

The classical equivalence between the `old' and `new' (S)CM models can hopefully be extended
to the quantum case and used to solve the quantum mechanics of the charged AdS$_2$
(super)particles in terms of (super)conformal quantum mechanics. In the classical hamiltonian
approach, this equivalence, both for the radial motion and with the angular $S^2$ variables
takeninto account, was proved in a recent paper \cite{000}.

\section{Conclusions}
In this talk a new kind of the relation between field theories possessing spontaneously broken
conformal symmetry in $d$-dimensional Minkowski space and the codimension-$(n+1)$ branes
in AdS$_{d+1}\times S^n$
type backgrounds in the static gauge was presented.
This relation takes place already at the classical level and transforms
the dilaton Goldstone field associated with the spontaneous breaking of scale invariance into the
transverse (or radial) brane coordinate completing the $d$-dimensional brane worldvolume to
the full AdS$_{d+1}$ manifold.
The conformally invariant minimal actions in Minkowski space including the dilaton are
transformed into nonlinear actions
given on the AdS brane worldvolume and involving, as their essential part, couplings
to the firsr extrinsic curvature
of the brane. Conversely, the standard worldvolume AdS brane effective actions prove
to be equivalent to some
non-polynomial conformally invariant actions in the Minkowski space.
The AdS/CFT map is one-to-one (at least, classically)
for the conformal actions containing no dilaton potential and for brane actions with
the vanishing
vacuum energy. The geometric origin of it can be revealed most clearly within the
nonlinear realization description of AdS branes \cite{dik} which generalizes
the analogous description of
branes in the flat backgrounds \cite{West}-\cite{ik}. In particular, it turns out that the
standard realization of
the conformal group in the Minkowski space and its transverse brane coordinate-dependent
realization
as the AdS$_{d+1}$ isometry group in the solvable-subgroup parametrization of AdS$_{d+1}$
are simply two alternative ways of presenting symmetry of the same system.
Most interesting subjects for further study are the generalization of the above
relationship to the case of AdS superbranes and, respectively, superconformal symmetries
in dimensions $d>1$, as well as the understanding of how it can be promoted to the quantum case.
Possible uses of the transformation \p{map} for the further analysis of relationships
between field theories on AdS$_{d+1}$ and ${\cal M}_d$ were already discussed in Sect. 4.

The existence of the coordinate map \p{map} suggests a novel view on the relationship between
the conformal field theory actions and the worldvolume actions of AdS superbranes.
As we saw, any conformal field theory action in the branch
with spontaneously broken conformal symmetry, after singling out the dilaton field, can be
rewritten in terms of the AdS brane variables, with the field-modified conformal
transformations defining the relevant symmetry. This relationship exists at any finite
and non-vanishing AdS radius
$R= 1/m\,$. It is interesting to further explore
this surprising `brane' representation of (super)conformal field theories,
especially in the quantum
domain, and to better understand the role of couplings to extrinsic curvature
which are unavoidable
in this representation. Let us recall that a string
with `rigidity', i.e. with the extrinsic curvature terms added to the action,
was considered as a candidate for the QCD string \cite{pol}.

\section*{Acknowledgments}
I thank S. Bellucci, S. Krivonos and J. Niederle in collaboration with whom
some of the presented results were obtained. I am grateful to the Organizers
of the Misha Saveliev Seminar for inviting me to give this talk and kind hospitality.
I thank the University of Hannover for warm hospitality at the final stage of this work.
A partial support from the grants INTAS 00-00254, DFG 436 RUS 113/669,
RFBR-DFG 02-02-04002, RFBR-CNRS 01-02-22005, RFBR 03-02-17440 and a grant of
Heisenberg-Landau Program is also acknowledged.

\end{document}